\documentclass{aa}
\usepackage{graphicx}
\newcommand{\acs}{$^{\prime\prime}$}

\newcommand{\mss}{mag arcsec$^{-2}$}

\begin{document}
\title{The z$\le$0.1 Surface Brightness Distribution} 
\author{K. O'Neil
\and S. Andreon
\and J.-C. Cuillandre}
\institute{NAIC/Arecibo Observatory; HC3 Box53995; Arecibo, PR 00612; USA
\and INAF-Osservatorio Astronomico di Capodimonte, via Moiariello 16, 80131 Napoli, Italy
\and Canada-France-Hawaii Telescope Corporation, P.O. Box 1597, Kamuela, HI 96743.}
\offprints{Karen O'Neil, \email{koneil@naic.edu}}

\date{Received / Accepted}

\abstract{
The surface brightness distribution (SBD) function describes the 
number density of galaxies as measured against their central
surface brightness.  Because detecting galaxies with low central surface
brightnesses is both time-consuming and complicated, determining the
shape of this distribution function can be difficult.
In a recent paper Cross, et al. suggested
a bell-shaped SBD disk-galaxy function which peaks near the canonical Freeman
value of 21.7 and then falls off significantly by 23.5 B mag arcsec$^{-2}$.  
This is in contradiction to previous studies which have typically
found flat (slope=0) SBD functions out to 24 -- 25 B mag arcsec$^{-2}$
(the survey limits).  Here we take advantage of a 
recent surface-brightness limited survey by Andreon \& Cuillandre which
reaches considerably fainter magnitudes than the Cross, et.al sample
(M$_B$ reaches fainter than $-$12 for Andreon \& Cuillandre while 
the Cross, et.al sample is limited to M$_B$ $<$ $-$16)
to re-evaluate both the SBD function as found by their data and the SBD
for a wide variety of galaxy surveys, including the Cross, et al. data. 
The result is a SBD function with a flat slope out through
the survey limits of 24.5 B mag arcsec$^{-2}$, with high confidence limits.
}

\maketitle

\section{Introduction}
The Surface Brightness Distribution (SBD) function --  a measure of the number
density of galaxies broken into bins of decreasing central surface brightness --
provides a quantitative description of the galaxy population within the Universe.
The first attempt at quantifying the local (z$\le$0.1) disk-galaxy SBD was done by \cite{freeman70}
who found a Gaussian distribution with $\langle \mu_B(0)\rangle$ = 21.65 $\pm$
0.30 mag arcsec$^{-2}$.  In the years since Freeman's distribution was published
a significant quantity of galaxies
have been found with central surface brightnesses more than 10$\sigma$ from
Freeman's canonical value, showing that Freeman's distribution clearly underestimated
the number of galaxies at faint surface brightness.   Indeed it is fairly certain 
that the distribution seen by Freeman was due to selection effects imposed by 
the considerable noise inherent in imaging techniques at the time, which effectively
eliminated the possibility of seeing galaxies with $\mu_B(0)\;\ge$ 23 mag arcsec$^{-2}$.

In the thirty years since Freeman's (1970) results were published, a number of
attempts have been made to describe the local SBD.  Adding on to the work done first 
by \cite{disney76} and then \cite{mcgaugh96}, \cite{oneil00} found that the SBD
of galaxies at $z<0.1$ is described by a curve which rises steeply
from 20 to 22 B mag arcsec$^{-2}$ and then remains flat through the survey limits
of 25.0 B mag arcsec$^{-2}$.  This implies that the
majority of galaxies in the local Universe are low in surface brightness, and
that LSB galaxies should play a significant role in studies of the local baryon 
density, damped Lyman-$\alpha$ systems, and in theories of galaxy formation and
evolution.

Recently, though, the belief in a flat SBD out to $\ge$ 25 mag arcsec$^{-2}$ has
been questioned.  Using a subsample of galaxies taken from the 2dF Galaxy
Redshift Survey, \cite{cross01} found the local SBD to be best represented by a
broadened version of Freeman's (1970) original SBD. If this is correct it would
have far-reaching implications.   First, a SBD which falls off before $\mu_B(0)$
= 24.0 mag arcsec$^{-2}$, as the Cross, et al. distribution does, would imply
that LSB galaxies are extremely rare and thus
are rarely the cause of phenomenon such as damped Lyman-$\alpha$
systems.  Additionally, though, the Cross, et al. distribution suggests that, as
surveys can now readily reach surface brightnesses fainter than 25.0 B mag
arcsec$^{-2}$, we have now seen the entire range of galaxies which exist at this
epoch.

As the \cite{cross01} results are both highly significant and 
seem in contradiction to the studies done by, e.g. \cite{oneil00}
and \cite{mcgaugh96}, further investigation is clearly warranted.
With this in mind we have taken the recent results from a Canada-France-Hawaii
Telescope (CFHT) deep survey of the Coma cluster (Andreon \& Cuillandre 2002) to
obtain an independent measure of the local SBD.  Our results are then
combined with previous SBD measurements and compared with that given by
Cross, et al.

\section{The Data}

B, V, and R observations of the Coma Cluster using the CFHT were obtained 
by \cite{andreon02}, and details about the observations and data reduction are
given within that reference.  All observations were taken on 12 January, 1999
with the Canada-France-Hawaii Telescope and CFH12K instrument
(Cuillandre, et al. 2002). The fields were centered on the Coma cluster and had a
usable area (observed field minus vignetting, etc.) of 0.29 degrees sq. in V
and R, and 0.20 degrees sq. in B. The total integration time was 720s for the
B and V images, and 480s for the R image.  The seeing was found to be
0.88\acs, 1.23\acs\ and 1.04\acs\ for the B, V, and R images, respectively.

For completeness, the sample was cut at central surface brightnesses 1.0 -- 1.5
mag arcsec$^{-2}$ brighter than the lowest detectable  objects.  That is, the 
sample is complete to the cutting central surface brightnesses of  23.75, 24.25,
and 23.75 mag arcsec$^{-2}$, where the limiting detection brightnesses are 
$\mu(0)$ = 25.0, 25.5, 24.5 mag arcsec$^{-2}$ in $B, V,$ and $R$, respectively.
At the cut-off limits the measured signal-to-noise ratio is $\sim$20.

Foreground and background
galaxies were eliminated from the sample by comparing counts of galaxies within
the observed field and within a control field which crosses the Coma supercluster.
Errors incurred in this method are discussed in detail in \cite{andreon02}
and are included in the error estimates for this data with the minor
difference that possible over--Possonian number galaxy fluctuations are not
taken into account for lack of knowledge on the fluctuation amplitudes.

As described in \cite{andreon02}, central surface
brightnesses were determined by finding the magnitude within
the 0.25 kpc aperture and dividing that by the area (in arcsec)
of that aperture.  (All analysis was done on images convolved with the
seeing disk.)  For the purpose of this
article, the galaxies were then separated into bins 0.5 mag arcsec$^{-2}$
wide, and counts were made to the number of objects in each bin.
The $R$ band image proved to be the most reliable, having a definitive
counts of $\sim$1000 galaxies (after elimination of background sources, etc.),
with bins containing between 16 -- 200 galaxies/bin in the $\mu_R(0)$ = 20 -- 23.5 
mag arcsec$^{-2}$ range.  The $V$ and $B$ images had total counts of 404 and 
157 galaxies, respectively, with bins containing 13 -- 26 galaxies/bin in 
$\mu_V(0)$ = 20 -- 24 mag arcsec$^{-2}$ and 8 -- 40 galaxies/bin in
$\mu_B(0)$ = 20.5 -- 23.5 mag arcsec$^{-2}$, respectively.  The considerably
higher counts in the $R$ band are due to a combination of lower sky noise, 
high CCD quantum efficiency, low galaxy background contamination,
and larger field size.  As a result, the
SBD determined for the $R$ band is by far the most reliable.

\begin{figure}
\resizebox{\hsize}{!}{
\includegraphics{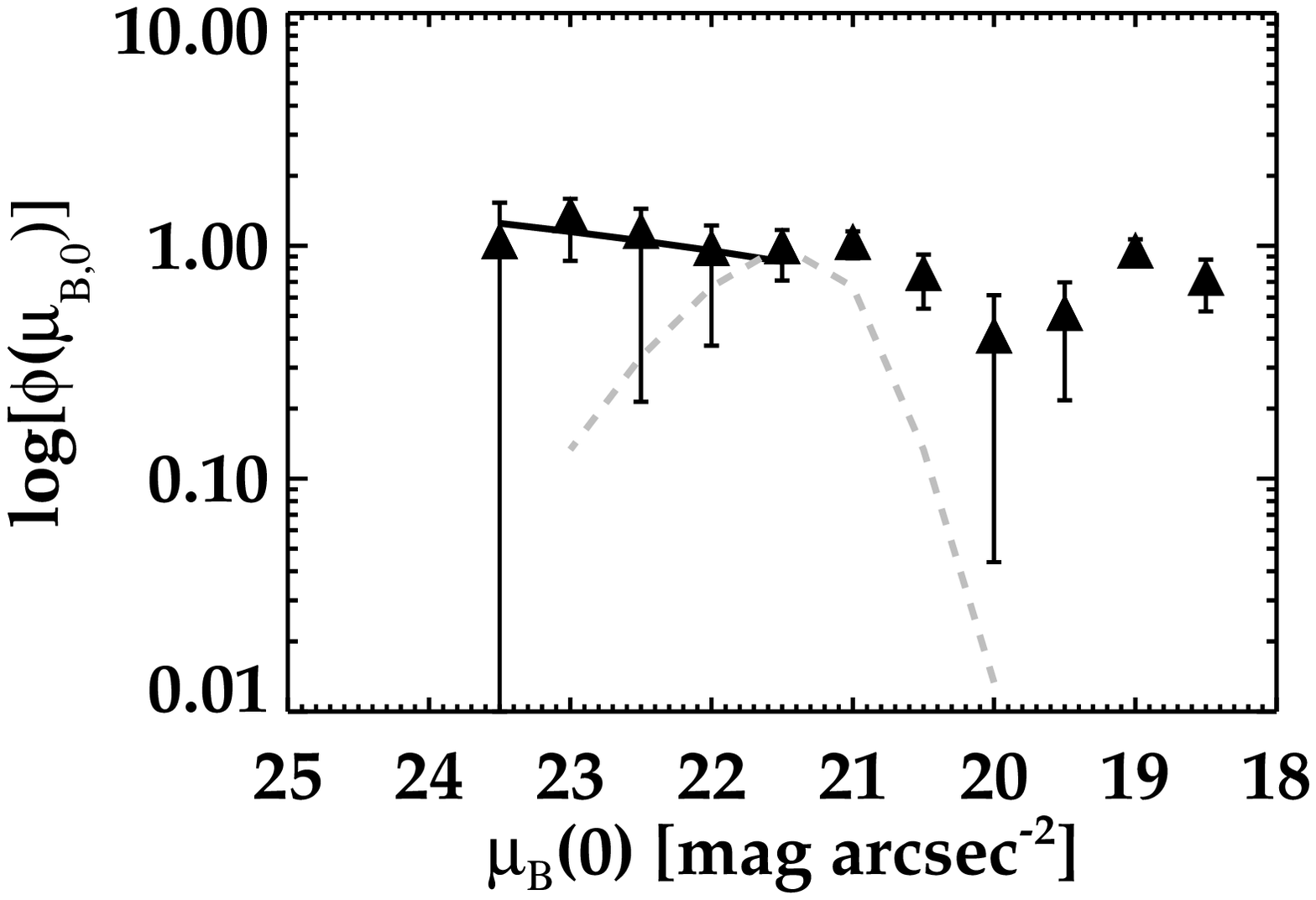}}
\caption{Plot of the data from \cite{andreon02}.  The best fit line for the $\mu_B(0)\;
\ge$ 21.5 mag arcsec$^{-2}$ data is given by the solid line, while the distribution of
\cite{cross01} is given by the dashed line.  The data has been normalized so that it
has a value of 1.0 at $\mu_B$(0)=21.6 mag arcsec$^{-2}$.  The Cross, et al. line cuts off
at 23.0 \mss, after which Cross, et al. state that a lack of data points makes their
curve unreliable.}
\label{fig:and_sample}
\end{figure}

\section{Finding the Surface Brightness Distribution}

\subsection{The Andreon \& Cuillandre Data}

Unlike many previous SBD studies (bar that of Cross, et al. 2001), the data from
\cite{andreon02} contains not just disk systems, but galaxies ranging from E
and S0 through pure disk systems.  As a result, at the bright end (e.g. between
18--21 B mag arcsec$^{-2}$) the sample is predominantly bulge-dominated 
(Figure~\ref{fig:and_sample}).  Fortunately, the contribution of bulge-dominated
galaxies at the surface brightness regime of interest (e.g. $\mu_B(0)\;\ge$ 21.5
mag arcsec$^{-2}$) is extremely small.  To account for this, all subsequent analysis of
the \cite{andreon02} data will take into account only those data points with
$\mu_B(0)\;\geq$ 21.5 mag arcsec$^{-2}$ or $\mu_R(0)\;\geq$ 20.0 mag
arcsec$^{-2}$.

\begin{figure*}
\resizebox{\hsize}{!}{
\includegraphics{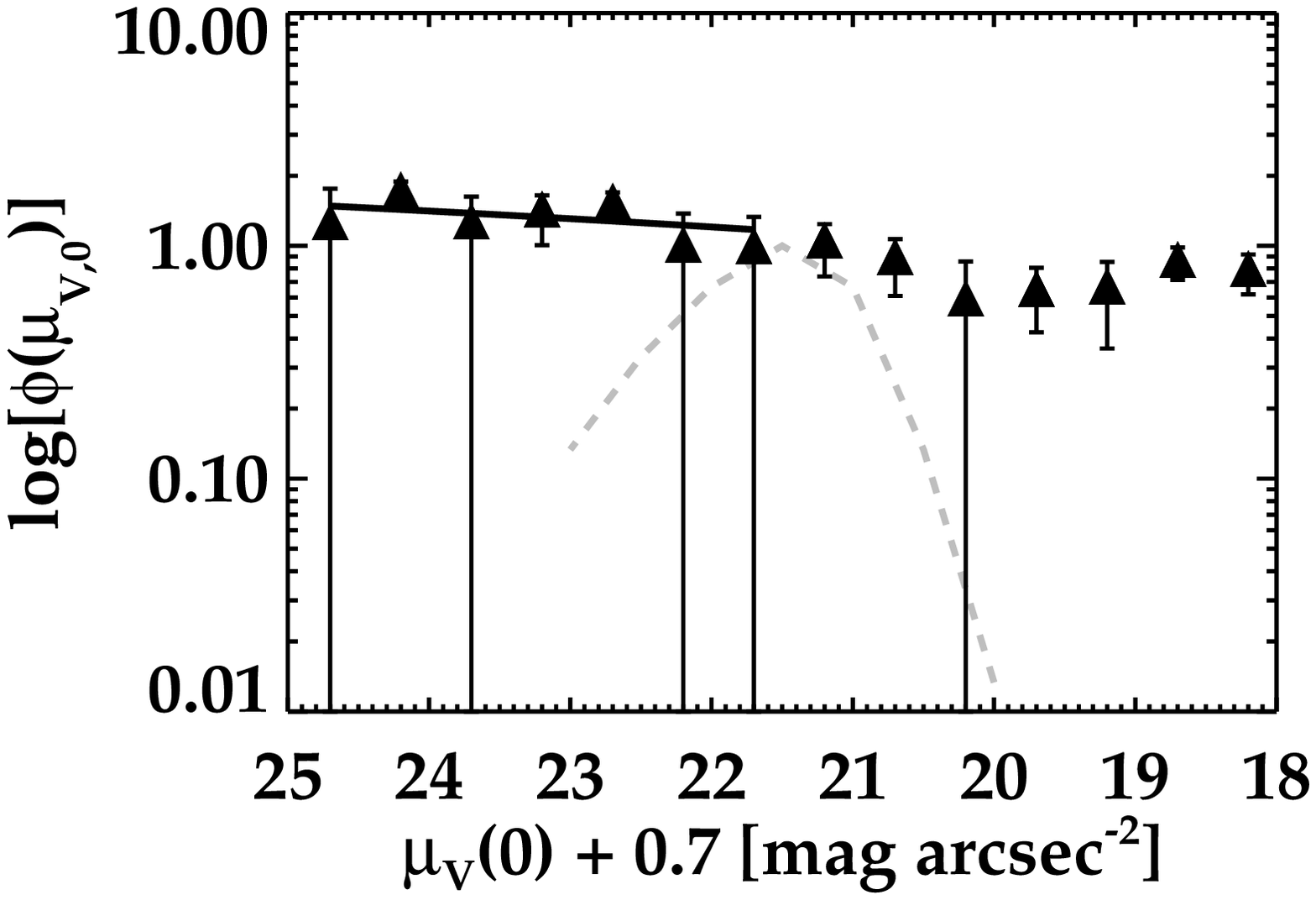}
\includegraphics{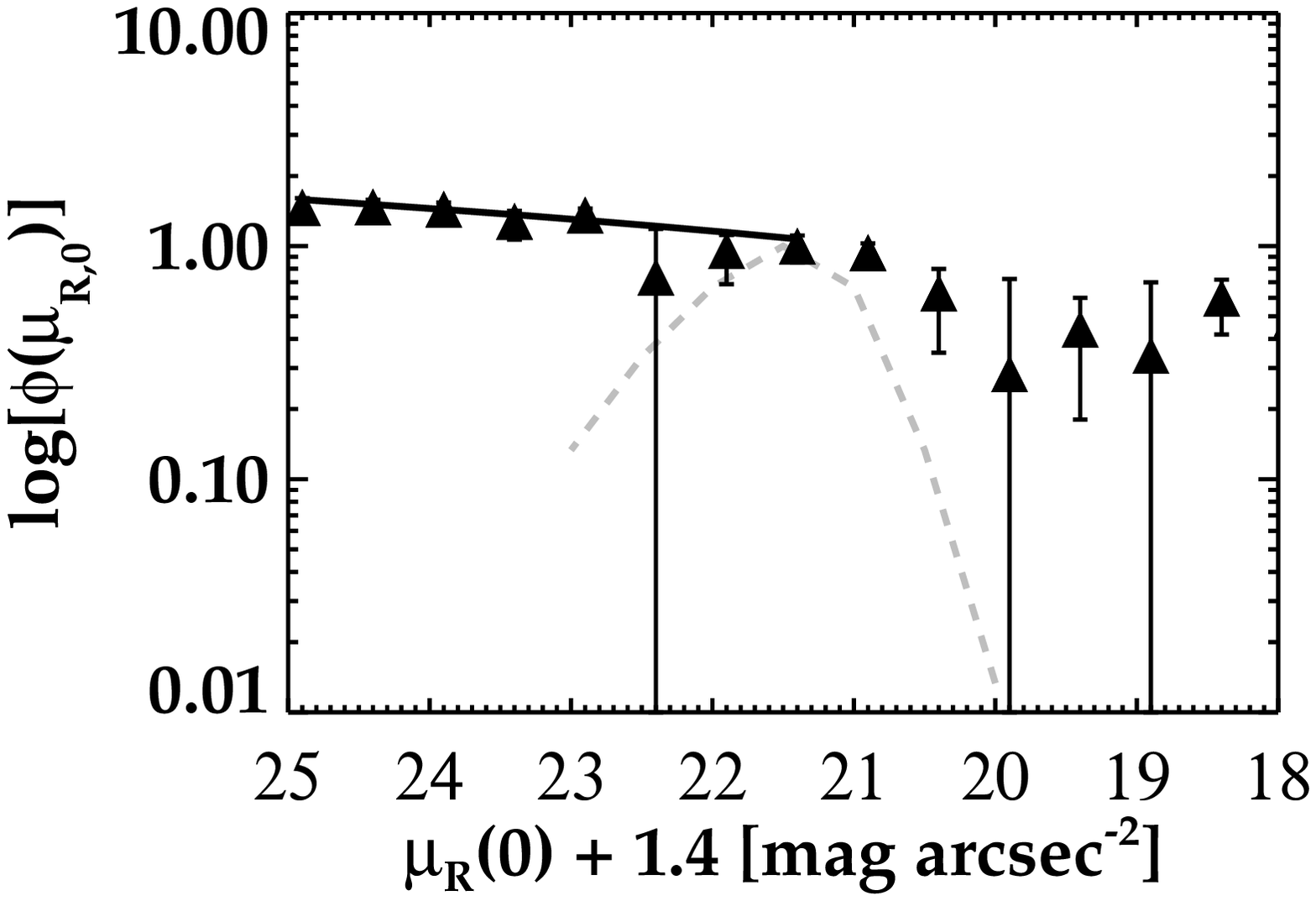}}
\caption{Plot of the V and R band data from \cite{andreon02}. Again, the distribution of
\cite{cross01} is given by the dashed line.  All data has been normalized so that it
has a value of 1.0 at $\mu_B$(0)=21.6 mag arcsec$^{-2}$.}
\label{fig:and_VR}
\end{figure*}

Obtaining a best-fit line to the $\mu_B(0)\;\geq$ 21.5 mag arcsec$^{-2}$ data
of  \cite{andreon02} gives a line whose slope is marginally {\it
increasing} with decreasing central surface brightness (slope = 0.08). This is
in marked difference to the results shown by \cite{cross01} which shows an
almost Gaussian distribution to the SBD (Figure~\ref{fig:and_sample}).  As the
error bars for the \cite{andreon02} distribution are fairly large, though, it
is conceivable that the Cross, et al.  distribution could describe Andreon \&
Cuillandre's $B$ band data.  

Strong support can be given to the argument that  Andreon \& Cuillandre's $B$
band data is best fit by a roughly horizontal line by looking at the SBD in $V$
and $R$ (the more reliable datasets) -- Figure~\ref{fig:and_VR}.   As discussed
above, the errors for Andreon \& Cuillandre's  V and R SBD are considerably
smaller than for their $B$ band data.  In particular, the $R$ band data has $>$
200 galaxies/bin in the lower ($\mu_R(0) \le 21.5$ mag arcsec$^{-2}$)
bins.   As can be seen, both the $V$ and $R$ band data have similar fits to the
$B$ band fit, giving strong credence to the argument that the Cross, et al.
curve is not an accurate description of the Andreon \& Cuillandre data.

\subsection{Combining Datasets}

\begin{figure}
\resizebox{\hsize}{!}{
\includegraphics{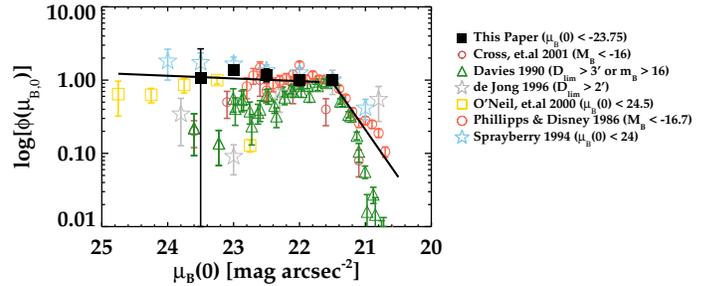}}
\caption{Plot of the data from a variety of galaxy surveys.  The associated surveys,
the survey limits, and the approximate magnitude limits of the surveys are given 
at the right.  The two lines
are the best-fit lines to all the data, broken into that data with $\mu_B(0)\;<$ 21.7 
mag arcsec$^{-2}$ and $\mu_B(0)\;\ge$ 21.7 mag arcsec$^{-2}$.
\label{fig:all}
}
\end{figure}

Another simple way to determine whether the \cite{cross01} distribution 
represents the true galaxy population at z$\le$0.1 is to combine all
previous data sets obtained for studying the local SBD function, and then obtain
a best-fit curve to the data.  Figure~\ref{fig:all} shows all data points used
for the SBD functions of \cite{mcgaugh96}, \cite{oneil00}, and \cite{cross01}
with the $\mu_B(0)\;\geq$ 21.5 mag arcsec$^{-2}$ points from \cite{andreon02}
overlaid. A best-fit line, weighted by the errors of the data points, is also
shown. To allow for the natural fall-off in central surface brightness at  
$\mu_B(0)\;\leq$ 21.7 mag arcsec$^{-2}$, two different lines were fit to
the data -- one for $\mu_B(0)\;\geq$ 21.7 mag arcsec$^{-2}$ and one for
$\mu_B(0)\;<$ 21.7 mag arcsec$^{-2}$.  In this case the  distribution again
shows a slight upwards slope at lower surface brightnesses
(slope=0.03).


Finally, perhaps the most accurate fit to the data is obtained by re-binning and statistically
averaging the data points from all previous studies, and obtaining a best-fit
line to this new data set.  In this case, the data was placed into 0.5 mag arcsec$^{-2}$ bins.
The mean (and error) were calculated, with each data point weighted inversely by its own variance.
The results of this re-binning are shown in Figure~\ref{fig:mean}.  Two best-fit
lines, again separated at the $\mu_B(0)\;=$ 21.7 mag arcsec$^{-2}$ mark, 
are shown.  In this case, the slope for the lower surface brightness regime 
has a slightly downward angle (slope=$-$0.1).
To insure that limiting the fit to a line did not force an artificial flattening
of the SBD function, we also attempted to fit a second order function to the data.
Not surprisingly, given the shape of the data, attempting to fit a bell-shaped curve 
to the seemingly flat data was unsuccessful.

\subsection{Magnitude and Measurement Differences}

One possible reason for the discrepancy between the Cross, et al.
(2001) SBD and that presented in this paper is that the 
selection criterion and method for determining $\mu(0)$ 
for the two surveys are quite different.  The Cross, et.al
survey uses an absolute magnitude cut-off at M$_B\;<\;-16$
and determines $\mu(0)$ by assuming an exponential profile and
extrapolating $\mu(0)$ from the measured isophotal magnitudes
and areas.  In contrast, the Andreon \& Cuillandre (2001)
sample uses a surface brightness limit and obtains all surface
brightness measurements through finding the total magnitude within
a central 0.25 kpc aperture and dividing that by the area
of that aperture (Section 2).
%
%
\begin{figure}
\resizebox{\hsize}{!}{
\includegraphics{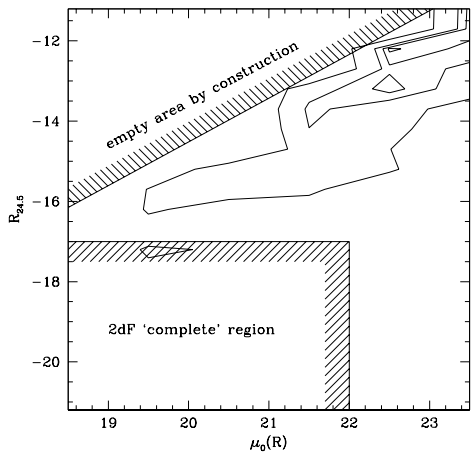}}
\caption{ 
Contours of the galaxies in the Andreon and Cuillandre (2002)
sample showing the number of galaxies in 0.5 R \mss\ and 1 R magnitude bins.
The first contour line indicates 10 galaxies and the lines are spaced by 40 galaxies/bin.
The enclosed area labeled `2dF complete region' demarcates the
completeness region for the Cross, et.al (2001) sample.  The region in the
upper left corner of the plot, labeled `empty area by construction'
demarcates the minimal galaxy size to be included in the the Andreon
\& Cuillandre sample.
\label{fig:comp}
}
\end{figure}

In the Coma sample, only 30 galaxies out of the 405 with $\mu_0(B)\;<\;$ 23.5 \mss\
also have M$_B$ $<$ $-$16.  As a result it is not practical 
try to mimic the Cross, et.al survey limits (Figure~\ref{fig:comp}). 
Instead, we can look at a variety of surveys undertaken
and the limits inherent in those surveys.  Figure~\ref{fig:all} shows the SBD for a
number of surveys, each having different survey limits.  As with the Andreon \& Cuillandre
data, the majority of these surveys do not have explicit 
magnitude limits but instead have a surface brightness and diameter limit.
We can, though, determine a rough magnitude limit for the surveys.  Having done
so (Figure~\ref{fig:all}), it is notable that no trend is clearly seen between the 
fall-off in the SBD and the magnitude limit of the survey.  This re-emphasizes the
idea that {\it any complete survey of galaxies is defined not by a limiting magnitude,
but by the combination of a limiting magnitude and diameter.}  That is, by
a surface brightness limit.

\section{Discussion}

The SBD described herein, obtained though combining a wide variety of
survey data (including that of Cross, et al. 2001), 
is in clear agreement with the studies of both \cite{mcgaugh96}
and \cite{oneil00}.  That is, up to the general survey limits
of 24.5 $B$ mag arcsec$^{-2}$, our data conclusively shows that the
SBD for the Andreon \& Cuillandre sample does {\it not} decrease
significantly between the canonical Freeman (1970)
value of 21.7 and the survey limits of 24.5 mag arcsec$^{-2}$.  Within the errors
of this data, the line can be best described as have a slope of 0.0 $\pm$ 0.1.
This is in clear contrast to the SBD determined by Cross, et al. (2001).
It is possible that the reason for the difference in the two
surveys is due to a much higher magnitude cut-off for the 
Cross, et al. sample than that of Andreon \& Cuillandre.
If this is correct it would imply that although low surface brightness
galaxies may numerically dominate the number counts of galaxies in the
local Universe, they do not play an important role in measures
of either the total light or mass at z $<$ 0.1.
It is important to note, though, that by combining a wide variety of survey data,
we attempted to minimize systematic errors induced by individual surveys and techniques.
That is, we have included both
samples which are purely volume limited (e.g. the Andreon \& Cuillandre sample)
and samples which are magnitude  and/or diameter limited (e.g. Sprayberry 1994) 
and no trend is clearly seen between the fall-off in the SBD and
the magnitude limit of the survey.

\begin{figure}
\resizebox{\hsize}{!}{
\includegraphics{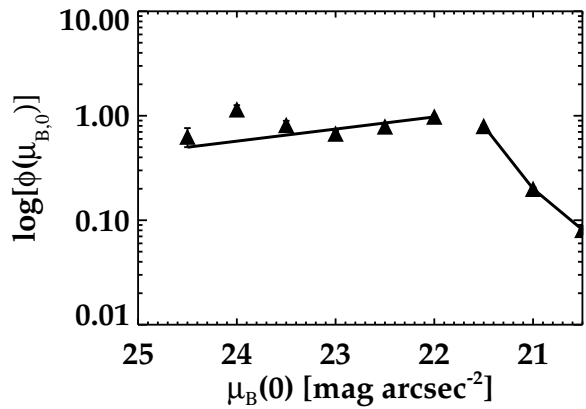}}
\caption{The same data as in Figure~\ref{fig:all} re-binned and averaged into 0.5 mag arcsec$^{-2}$
bins.  The best fit lines to this data are shown, where again the lines are separated at
the $\mu_B(0)\;=$ 21.7 mag arcsec$^{-2}$ mark.}
\label{fig:mean}
\end{figure}

It seems plausible that the SBD given in Figure~\ref{fig:mean}
is an accurate representation of the SBD in the z $<$ 0.2 Universe.
Due to a dearth of data, though, the shape of the SBD in the
$\mu_B(0)\;>$ 25.5 mag arcsec$^{-2}$ range cannot currently be determined,
as finding the true form of the SBD at lower central surface brightnesses
will only happen as survey sensitivities increase.
We can gain a hint of what be be found through examining some recent findings in the
literature.  There have been a number of discoveries over the
past few years of galaxies detected at 21-cm which which cannot be identified
down to optical limits of 25$-$27 mag arcsec$^{-2}$ 
(Boyce, et al. 2001; Ryder, et al. 2001; Kilborn, et al. 2000; Rosenberg \& Schneider 2000).
Although none of these detections can make any statement
as to the number density of extremely  low surface brightness galaxies, the
fact that any galaxies have been found so far below current survey limits argues
that we still are not seeing a complete picture of the local galaxy population.

\end{document}